%Paper: 9204017
%From: atish@physics.rutgers.edu
%Date: Wed, 29 Apr 92 16:03:54 EDT

\input harvmac
\Title{\vbox{\baselineskip12pt\hbox{RU-92-09}
\hbox{hep-lat/9204017}}}
{\vbox{\centerline{Decoupling a Fermion Whose Mass}
\centerline{Comes From a Yukawa Coupling:}
\centerline{Nonperturbative Considerations}}}

\centerline{T. Banks$^1$, A. Dabholkar$^2$}
\baselineskip18pt
\centerline{Dept. of Physics and Astronomy}
\centerline{Rutgers University}
\centerline{Piscataway, NJ 08855-0849}
\smallskip
\baselineskip18pt
\noindent
Perturbative analyses seem to suggest that fermions whose mass comes
solely from a Yukawa coupling to a scalar field can be made arbitrarily
heavy, while the scalar remains light.  The effects of the fermion can
be summarized by a local effective Lagrangian for the light degrees of
freedom.
Using weak coupling and large N
techniques, we present a variety of models in which this conclusion is
shown to be false when nonperturbative variations of the scalar field
are considered.  The heavy fermions contribute nonlocal terms to the
effective action for light degrees of freedom.  This resolves paradoxes
about anomalous and nonanomalous symmetry violation in these models.
Application of these results to lattice gauge theory imply that
attempts to decouple lattice fermion doubles by the method of Swift and
Smit cannot succeed, a result already suggested by lattice calculations.

\vskip 1cm
\noindent

\footnote{}{$^1$ (banks@physics.rutgers.edu)}
\footnote{}{$^2$ (atish@physics.rutgers.edu)}

\Date{April 1992}
%\draft

\newsec{Prelude and Paradox}

As its title suggests, this paper should be thought of as a
continuation of the work of  D'Hoker and Farhi
\ref\df{
E. D'Hoker, E. Farhi, {\it Nucl. Phys.}{\bf B248}, (1984), 59 \& 77.}
on the decoupling of
heavy fermions which transform in chiral representations of a
spontaneously broken gauge group.  This is a phenomenon which is crucial
to several ideas in modern particle physics.  The conventional wisdom
holds that the chirality of the observed fermion representations is a
fundamental property of the world, but mirror fermions could be
discovered at the next accelerator.  Are there theoretical bounds on how
large their masses could be?  If indeed chirality is fundamental and not
a low energy accident it may have profound implications, for we know of
no gauge invariant regulator for chiral gauge theories, nor any real
argument that they are consistent outside the realm of perturbation
theory.\foot{Among the many pleasing aspects of string theory is the
natural cutoff it provides for theories of chiral fermions. D. Friedan
has argued that this should be viewed as a hint that string theory
rather than pointlike field theory describes the real world.  Note that
although string theory does not yet give a nonpertubative description of
chiral gauge theories, it is a finite gauge invariant regulator in
perturbation theory.  All other perturbative regulators break chiral
gauge invariance explicitly.}
Attempts to construct chiral gauge theories as continuum limits of
honest lattice field theories with short range couplings are hampered by
the Nielsen-Ninomiya theorem.  In the naive lattice version of the
standard model, this
theorem guarantees the existence of mirror
partners of quarks and leptons in the continuum limit.  One can only
hope to decouple them by giving them large Yukawa couplings to the Higgs
field, and perhaps masses of the order of the cutoff. The success of
this program would imply that there can be no theoretical upper bounds
on the masses of mirror fermions.  If lattice gauge theorists can send
them off to infinity on the computer, God should be able to do the same
in the real world.  In order to argue against the existence of very
heavy mirror fermions one would be reduced to complaining about fine
tuning (the question of how much work we believe God is willing to do)
or the failure of perturbation theory (the question of how much work we
are willing to do).

In a recent paper\ref\banks{T. Banks, {\it Phys.Lett.}{\bf B272},(1991),75. },
one of the authors pointed out a possible problem
with most attempts to construct the lattice standard model by these
techniques.  Any $SU(3)$ lattice gauge theory with no colored Higgs
fields and a Lagrangian bilinear in fermions has a global $U(1)$
symmetry that acts on the lattice quark fields like baryon
number.\foot{In almost all theories, this symmetry can be gauged on the
lattice.  The single exception of which we are aware is a theory in
which $\bf 3$ and ${\bf{\bar 3}}$ fields are put on different sites of an
Euclidean lattice.  This theory has a global baryon number symmetry which
cannot be gauged. It is an interesting example of a lattice gauge theory
where the Lagrangian is gauge invariant but the functional measure
isn't. }
This conserved baryon number symmetry would appear to forbid
the nonperturbative baryon number violating process discovered by 't
Hooft\ref\hofft{G. 't Hooft,{\it Phys.Rev.Lett.}{\bf 37},(1976),8.}
in the semiclassical approximation to the continuum standard model.
If baryons are constructed from quark operators in any quasilocal way,
lattice Green's functions with non zero baryon number will vanish
identically for all values of the parameters on the lattice.\foot{We are
assuming that in those cases where equal numbers of ${\bf 3}$ and
${\bf{\bar 3}}$ fields sit on each site,
the U(1) symmetry is not spontaneously broken.
We believe this to be the conventional wisdom. If it did suffer
spontaneous breakdown the theory would contain a Goldstone boson not
observed in nature, and would not converge to the standard model.}
Thus, either the lattice theory does not succeed in reproducing the
conventional continuum model in perturbation theory, or we have
discovered a nonperturbative violation of universality.

In an attempt to understand this puzzle without resorting to a computer,
we have constructed a model in which heavy fermion decoupling can be
studied entirely within the framework of weakly coupled continuum field
theory.  The model was motivated by a 1988 remark of David Kaplan. In
order to turn a vectorlike gauge theory into one with a large hierarchy
between fermions and their mirror partners, we must make the Yukawa
couplings of the mirrors to the Higgs fields which break the gauge
symmetry much larger than the gauge coupling.  Kaplan remarked that this
was possible within the perturbative domain, if we are willing to
consider theories with extremely small gauge coupling.  While not
directly applicable to the real world, such models might prove to be
an interesting theoretical laboratory.  This is indeed the case as we
will see below.

Our model then begins as the standard model with all of the usual
couplings scaled down by a factor $f$.  For definiteness we might
consider $f\sim 10^{-2}$.  We add to this a set of mirror fermions, left
handed Weyl fields that transform in the complex conjugate
representation of the standard model fermions.  The vacuum expectation
value of the Higgs field is arranged to be $\sim 250$ GeV as usual, and
the mirror fermions are all given Yukawa couplings $g_{mirror}$
of order one, so that
their masses are of order $100$ GeV.  The gauge bosons and conventional
fermions have masses below $1$ GeV. We forbid the gauge invariant mass
terms that could be made by pairing conventional fermions with their
mirror partners.  This is natural, due to a symmetry that will be
discussed below.
The hierarchy between vector boson and heavy fermion masses in this
model requires no fine tuning.  Radiative corrections to the squares of
gauge boson
masses due to loops of mirror fermions are of order ${e_{gauge}^2
g_{mirror}^2 \over 4\pi^2} v^2$, and are small compared to the tree
level masses.  Note however, that if we insist that the Higgs boson
mass be as small as the vector boson mass, the conventional vacuum state
becomes metastable.  This is a consequence of the familiar unboundedness
of the fermionic one loop correction to the effective potential.  In the
present model, when the Higgs and vector boson masses are a hundred
times smaller than the fermion mass, the turnover of the effective
potential occurs in a region accessible to perturbation theory and one
might think that the conclusion that the vacuum is only metastable is
reliable.  If this is the case, then
 the discussion below can be read as a description of processes
going on in this metastable state, and one is confronted with issues of
the relative rates of the 't Hooft process and the decay of the false
vacuum.  We note however that Kuti and Shen\ref\kuti{Y.Shen, in {\it
Conference Proceedings of Lattice 90}, Tallahassee, Florida, {\it Nucl.
Phys.} {\bf B20} (Proc. Suppl.),(1991),613; J.Kuti, Y. Shen,
UCSD-preprint UCSD/PTH 91-18.} have argued that in a theory with only
bare quartic couplings one cannot attain the renormalized parameter
values for which the vacuum is metastable.  The Higgs mass remains a
finite fraction of the fermion mass for all parameter values.  We do not
know if this conclusion remains true in the presence of irrelevant
couplings in the bare Lagrangian, or when the system is coupled to gauge
fields.

We believe that the issue of metastability of the perturbative vacuum as
we vary the relative ratio of fermion to Higgs
masses is a crucial one, and we will have much more to say about it in
 section 3 when we examine a two dimensional model in the large
$N$ limit.  There we will show that by fine tuning of many parameters
we can obtain a model with a stable symmetry breaking vacuum
in which the ratio of the fermion mass to both
vector boson and Higgs boson masses is extremely large.  The puzzle we
describe in the next paragraph exists in that model as well.  Therefore
we ask the reader to ignore issues of vacuum metastability for the moment.

A bit of thought about nonperturbative baryon number violating processes
in this model reveals an apparent paradox, whose resolution will be the
subject of this paper.  The baryon number current built out of mirror
quarks, has an $SU(2)$ gauge anomaly which is exactly equal to that of
the ordinary baryon number current.  Thus, the difference between
ordinary and mirror baryon numbers is an exactly conserved anomaly free
symmetry.  Coupled with the fact that all mirror baryons have masses of
order $100$ GeV, this symmetry forbids the decay of particles with
ordinary baryon number and masses of order $1$ GeV or below, since any
such decay would have to produce mirror baryon number and there are no
light particles that carry this quantum number.\foot{Note
that a very similar argument appears in 't Hooft's original calculation
of deuteron decay in the standard model.  If first and second generation
baryon numbers were separately conserved, the deuteron could not decay.
Its decay rate vanishes with the Cabibbo angle.  By omitting the mass
term mixing ordinary and mirror fermions we have eliminated the
corresponding \lq\lq Cabibbo mixing'' in our model.}

Now let us study the same model using conventional ideas of decoupling
and low energy effective field theory.  The particle spectrum at $1$ GeV
and below coincides with that of the conventional standard model
rescaled by $f$.  One might conclude then that the physics at this
energy scale was well described by the standard model with rescaled
couplings.  But then, 't Hooft's calculation of the deuteron decay rate
could be carried out, giving a result many orders of magnitude below
that in the standard model, but still nonzero!  This is in blatant
contradiction with
 the exact result demonstrated in the previous paragraph.
Note the similarity to the lattice models discussed in\banks.  The role
of mirror fermions is played by lattice doubles of the continuum
fermions. The U(1) symmetry discussed above is continuum baryon number
plus double mode fermion number.  If the doubles indeed have masses of
order the cutoff while the continuum fermions have their observed
masses, then we have a paradox very similar to that in the superweakly
coupled standard model.

The arguments of D'Hoker and Farhi\df do not seem to shed much additional
light on this situation.  These authors work in the limit of a fixed
length Higgs field. They tell us that if we try to compute the
mirror baryon number current in the low energy theory then it will be
equal to the Skyrmion number current of the nonlinear model representing
the unphysical Higgs degrees of freedom.  When the model is gauged,
this is the
same as the Chern Simons current built out of the gauge invariant
massive gauge fields.
It is a gauge invariant current whose divergence is proportional to the
$SU(2)$ topological charge.  The difference between it and the ordinary
baryon number current is an anomaly free gauge invariant conserved
current.  However, this gauge invariant Chern Simons current exists in
the standard model even when there are no mirror fermions.  Arguments
based on it cannot resolve our paradox unless they imply that 't Hooft's
calculation is wrong in the unextended standard model.  There is always
a gauge invariant conserved current which acts as baryon number when
applied to quark fields.  Furthermore, the change in Chern
Simons charge built from massive gauge fields is the integral of a total
derivative of a gauge invariant object constructed out of massive fields.  One
would expect it to be zero.  If the Chern Simons current were really a
well defined operator in the conventional standard model, this argument
would rule out baryon number violation completely.\foot{In section 3 we
will present a two dimensional model in which the D'Hoker Farhi
scenario is realized.  It is indeed the case that
baryon number is not violated in the low energy
effective action of this model.}  Thus, if we believe that 't Hooft's
calculation is correct in the unextended standard model, the arguments
of D'Hoker and Farhi cannot help us to understand the paradoxes of
decoupling in the model supplemented with heavy mirror fermions.

This is perhaps the place to discuss the criticisms of the arguments
of\banks made by Dugan and Manohar\ref\manohar{M.J. Dugan,A.V.
Manohar,{\it
 Phys.Lett.}{\bf B265},(1991),137. }.  These authors claim
to show that the conserved lattice current corresponding to the symmetry
discussed in\banks is not gauge invariant.  As we have stated it, this claim is
obviously wrong on the lattice.  We can define the current by gauging
the U(1) lattice baryon number discussed above (and varying with respect
to the background gauge field) and since every term in
the lagrangian is invariant under the standard model group, so is the
current.  As the authors of\manohar point out in their equation [14],
the real meaning of their calculation in a model in which gauge
invariance of the Wilson term is enforced by introducing a Higgs field,
is that the conserved current differs from the light baryon number
current by the Chern-Simons term of the massive gauge fields.\foot{Dugan
and Manohar are clearly working in the fixed length Higgs model, or
ignoring zeroes of the Higgs field.}  Thus their conclusions are
identical to those of D'Hoker and Farhi and do not really shed any more
light on the baryon number paradox.

We will present the resolution of this paradox in the next section.  It
is, we believe, rather surprising, and shows that the decoupling of a
fermion whose mass comes from a Yukawa coupling is profoundly different
than ordinary decoupling, even more so than one would have concluded
from the work of\df or from recent work on new parameters arising from
loops of heavy chiral fermions in electroweak radiative corrections.
In effect, what we will show is that although the
{\it particles} associated with mirror fields are heavy, the mirror
fields themselves do not decouple from low energy physics, as long as
the Higgs field is light.  Depending on
the configuration of low energy gauge and Higgs fields, an arbitrarily large
number
of modes of the mirror fields can contribute significantly to low energy
tunneling processes.  They completely transform the instanton dynamics
of the low energy gauge system.

As a counterexample to the claim that one can entirely decouple mirror
fermions, this weakly coupled model is not completely satisfactory.  The non
perturbative effects which exhibit this dramatic violation of decoupling
are, in the weak coupling regime, much smaller than the perturbative
effects of nonrenormalizable operators in the baryon number conserving
sector.  Thus there is not a completely clean separation of scales.  We
cannot reduce the perturbative effects of the heavy mirror particles to
arbitrarily small size
without leaving the realm of perturbation theory.  In addition, and more
importantly, if we try to make the Higgs mass much smaller than the
fermion mass in this model we are confronted with vacuum instability.\foot{
We will see later that this problem of vacuum instability is the real
iceberg on which decoupling founders, and that our paradox about baryon
number is only the tip of it.}
Nonetheless the fact
that the zero modes and lack of decoupling are evident for all values
of the mass that are amenable to a perturbative analysis, suggests that
the phenomenon that we have uncovered persists into the strong coupling
regime.  Even if we are able to construct a model with heavy fermions,
light Higgs, and a stable vacuum, we will still find that the fields of
the heavy fermions do not decouple from low energy physics.

To obtain further evidence for this, we examine in section 3 some two
dimensional chiral gauge theories which are almost soluble.  We show
that in a model with a fixed length Higgs field (which is renormalizable
in two dimensions), we can indeed decouple heavy mirror fermions.  Our
paradox about baryon number conservation is resolved by showing that
baryon number violating amplitudes vanish in the limit of fixed length
Higgs field.
When the modulus of the Higgs field is allowed to fluctuate this is not
the case.  We study the fluctuating length theory in the large $N$
approximation. In order to keep the radial mode of the Higgs field
light, and the classical vacuum stable,
we have to fine tune a number of parameters that grows with the fermion
mass. This is a consequence of a general property of decoupling of heavy
particles (gauge bosons as well as fermions) whose mass comes solely from
the vacuum expectation value of a scalar field.  It is quite generally
true that the
effective potential for the scalar induced by virtual heavy particles
is large and has
curvature of order the masses of the heavy particles.  It is also
nonanalytic when the Higgs field VEV goes to zero, because in this limit
the \lq\lq heavy'' particles become massless and the theory contains
infrared divergences.  Thus, if we perform no fine tuning, the mass of
the Higgs particle itself (the excitation of the radial mode of the
Higgs field) is large.  Further, because of the nonanalyticity of this
potential at the origin, we cannot fine tune the coefficients of
a finite number of analytic functions in the tree level potential to
cancel off the large effects of the heavy particles.  We will argue
below that the Linde-Weinberg lower bound on the Higgs boson mass
\ref\linde{A. Linde, {\it Sov. Phys. JETP Lett} {\bf 23} (1976) 64;
S. Weinberg, {\it Phys. Rev. Lett. } {\bf 36} (1976) 294.}
is another example of this effect.  It can be
viewed as a failure of decoupling of heavy vector bosons from an erstwhile
effective field theory for light Higgs bosons.
We believe that this fundamental obstacle to obtaining a light Higgs in
the presence of heavy fermions or bosons whose mass is driven by the
Higgs VEV is the real reason for the failure of decoupling of chiral
fermions.  We can obtain a model with a large fermion to scalar mass
ratio and a stable vacuum only by fine tuning many parameters.

 If, in our two dimensional model, we perform the infinite parameter
 fine tuning required to obtain a light Higgs and a stable vacuum, we
 still find problems. Heavy fermions have
light modes and do not decouple in the presence of configurations where
the
Higgs
field goes to zero in some regions of spacetime.  The effect of these
light modes is to drastically change the nonperturbative (in $N$)
physics of the low energy theory.  Baryon number violation and
confinement of fractional charges, which are both present in the model
without heavy fermions, disappear in the model with heavy fermions.

To summarize,
it appears very difficult to construct a model in which fermions that
get their mass from a Yukawa coupling to a scalar field are allowed to
have masses much larger than that of the mode which controls
fluctuations in the magnitude of the scalar.  In two dimensions, using
the infinite number of relevant operators at the scalar Gaussian fixed
point, it is possible to construct such models.  However, when the
system is coupled to a gauge field, there are light modes of the heavy
fermions in instanton configurations in which the magnitude of the
scalar field vanishes locally at certain points in spacetime.  These
light modes completely change the dynamics of the low energy theory.
The only way to truly decouple the fermions is to freeze the magnitude
of the scalar field simultaneously.  In this limit, instanton processes
have zero amplitude because the instanton action goes to infinity.
Thus all paradoxes related to chiral fermion decoupling are removed, but
at the price of \lq\lq throwing the baby away with the bathwater''.

In four dimensions, it seems highly unlikely to us that it is possible
to do the fine tuning necessary to keep the Higgs field light in the
presence of extremely massive chiral fermions.  The effective potential
generated by the heavy fermions naturally has an energy scale of the
fermion mass.  Furthermore it is singular at the origin of field space
and cannot be well approximated by a quartic polynomial.
Renormalizability restricts us to quartic polynomials, so we cannot
cancel the effect of the fermions with local counterterms.
The Higgs mass would be driven to infinity with the fermion mass.
Since there are no sensible continuum
theories with fixed length Higgs fields in four
dimensions\ref\dashneuberg{
 H. Neuberger,
LATTICE HIGGS AND YUKAWA MODELS,
Invited talk given at Int. Workshop Lattice '89, Capri, Italy, Sep
18-21, 1989.Published in Capri,Lattice 1989, 17
 R. Dashen, H. Neuberger,{\it Phys.Rev.Lett.}{\bf 50},(1983),1897.
},
this argument
suggests that it will be impossible to find a four dimensional model
with decoupled chiral fermions, and hence impossible to build a lattice
version of the standard model with many of the current local algorithms.
In any
case, no model built in this way can contain the 't Hooft mechanism for
baryon number violation.  If finely tuned models with light Higgs exist,
baryon number conservation will be enforced by confinement of instantons
through heavy fermion zero modes, while in models with fixed length Higgs
fields, instantons will have infinite action.  In the penultimate
section of this paper we will give a brief survey of attempts to
construct lattice standard
models and point out those which may evade the difficulties discussed in
this paper.

A disturbing possibility raised by our analysis of finely tuned models
is the occurence of important low energy fields which create only very
heavy particle states from the vacuum.  This dramatic failure of the
association between fields and experimentally accessible particle states
would make it difficult to find experimental tests of a theory
containing such {\it phantom} fields.
 The large $N$ model of section 3
certainly contains phantom fields.  One is led to ask whether their occurrence
is likely in the real world.  D'Hoker and Farhi\df suggested the existence
of fermionic solitons in the effective action generated by decoupled
chiral fermions.  These had the same quantum numbers as the original
fermions and masses of the order of the low energy scale.  The solitons
of D'Hoker and Farhi are topological excitations in a theory with fixed
length Higgs fields.  A related phenomenon\foot{We do not really
understand the relation between these two types of soliton.} is the
existence of baglike\ref\bag{P.Vinciarelli, {\it Lett. Nuovo
Cimento} {\bf 4} (1972) 905;  T.D. Lee, G.C.Wick, {\it Phys. Rev.} {\bf
D9} (1974) 2291; R.Dashen, B.Hasslacher, A.Neveu, {\it Phys. Rev.} {\bf
D10} (1974) 4114 \& 4130; W.Bardeen, M.
Chanowitz, S.Drell, M.Weinstein, T.M.Yan, {\it Phys. Rev.} {\bf
D11} (1975) 1094; R.Friedberg, T.D.Lee, {\it Phys. Rev.} {\bf
D15} (1977) 1694; {\bf D16} (1977) 1098; {\bf
D18} (1978) 2623; J.Bagger, S.Naculich, Johns Hopkins U.,
JHU-TIPAC-910018; {Phys. Rev. Lett.} {\bf 67} (1991) 2252.}
nontopological solitons in models with
a Higgs field of fluctuating magnitude.  In these configurations, light
states with single fermion quantum numbers are created by deforming the
Higgs field from its VEV over a finite region of space.  Since the
fermion mass is zero in the region where the Higgs field vanishes, these
states can be much lighter than fermions propagating in the vacuum if
the energy required to deform the Higgs is small compared to the fermion
mass.

Bagger and Naculich have recently studied these baglike solutions
in a strongly coupled large $N$ model\bag\foot{At large $N$,
as we will see in section 3, single fermion bags cannot form.  Bagger
and Naculich study bags containing $N$ fermions.}.  They find that these
states have mass comparable to the fermion mass in the strong coupling
region. However, they do not perform the fine tunings necessary to keep
the Higgs mass finite as the fermion mass goes to infinity (their model
is four dimensional, and it may not be possible to do this in a
consistent way).  Thus, it is not surprising that the bag picture, which
depends on an easily deformable Higgs field, fails in their model.  It
seems plausible however, that baglike solitons with single fermion
quantum numbers will exist in most models in which it is possible to
fine tune the Higgs mass to be much smaller than the fermion mass,
without destabilizing the vacuum.  These are precisely the models in
which one might suspect the occurrence of phantom fields.  The
existence of light bags in such models would eliminate the phenomenon of
phantom fields.  The phantoms would be interpolating fields for the
light bag states, and we could ascribe the nonperturbative dynamics
associated with them to the action of these particles.
Our two
dimensional large $N$ model is an explicit counterexample to the
conjectured existence of light bags in all models with phantom fields.
We will argue however that this may be a peculiarity of the large $N$ limit.

We have not really studied the question of the existence of light bag
states in much detail.  It deserves more attention, for it may be the
key to finding a theoretical upper bound on the mass of mirror fermions
or other as yet unobserved chiral representations of the standard model
gauge group.

\newsec{Massless Modes of Massive Particles}

Let us then study the weakly coupled version of the standard model
introduced in the previous section, ignoring questions of stability of
the perturbative vacuum.  That is, we will study classical solutions to
the Euclidean equations of motion, and the fermion determinant in these
backgrounds.    The crux of our argument is that the Euclidean
Dirac equation for mirror
fermions (or ordinary fermions for that matter)
in the standard model has such zero modes in the presence of an
instanton field.  Indeed, if we set the Yukawa couplings to the Higgs
field to zero, the existence of such modes is a trivial consequence of
the anomaly equations for mirror baryon number and lepton number.
Since the zero modes carry baryon number and the Yukawa couplings
preserve baryon number, there is no way for the Yukawa couplings to
lift these modes to nonzero (Euclidean) energy.

More mathematically, near the center of the instanton, the Higgs field
goes to zero and the gauge field approaches that of the instanton
solution of pure gauge theory.  The solution of the zero eigenvalue
Dirac equation in this region is
\eqn\zeroa{\psi_L = \psi_0 [A]}
\eqn\zerob{\psi_R = \psi_R^0}
where $\psi_0 [A]$ is the zero mode solution of the left handed Weyl
equation in the pure gauge instanton background, and $\psi_R^0$ is the
solution of the right handed Weyl equation with a source given by the
product of the Higgs field and $\psi_0 [A]$.  Since $\psi_0 [A]$ is not
singular at the origin, and the Higgs field goes to zero there, no
special choice of boundary conditions must be made to make the full
solution normalizable at the origin.  At infinity, the Higgs field goes
to a constant and the gauge field falls off exponentially (in unitary
gauge). The Dirac equation becomes that for free massive fermions.
There are exponentially increasing as well as exponentially decreasing
solutions of this equation, but since we have not used up any parameters
making the solutions regular at the origin, we have enough parameters
left to eliminate the exponentially increasing solution.  Consequently,
the zero modes are normalizable despite the fact that asymptotically the
fermion fields behave as if they were massive.

The existence of these zero modes means that amplitudes which involve
a change of topological charge, and involve only particles which exist
in the low energy theory, vanish identically.  The 't Hooft interaction
which describes the effect of instantons on the fermions in the theory,
is an operator which changes mirror baryon number. Its form is
\eqn\thooftlag{{\cal L}_{{\rm 't Hooft}} = \prod \psi_L \prod \psi_H}
where the products run over light and heavy fermion zero modes.
This interaction connects the
heavy sector to the light sector, but has no matrix elements within the
light sector itself.  Note that this is an exact consequence of the full
theory, but cannot be derived from a low energy Lagrangian from which
the mirror fields are omitted.  Thus, 't Hooft's calculation of baryon
number violation is radically altered in the theory with heavy mirror
particles.  It no longer predicts baryon number violation in the light sector.

It is worth pointing out that the dramatic violation of decoupling that
we have just discussed is actually implicit in 't Hooft's original
calculation of baryon number violation in the standard model.  't
Hooft included two generations of quarks and leptons in his calculation
of deuteron decay.  The
second generation quarks and leptons have instanton zero modes, and if
there is no Cabibbo mixing to convert these modes into modes of first
generation fermions, the amplitude for deuteron decay vanishes.  It is
proportional to $sin^3 \theta_{Cabibbo}$.  This by the way is the reason
that the deuteron rather than the proton decays by the 't Hooft process.
The instanton violates first generation baryon number by one unit, and
second generation baryon number by one unit, preserving their
difference. Cabbibo mixing violates individual generation baryon numbers
by $1\over 3$, preserving their sum.  The final change in baryon number
in a process in which no second generation particles is involved is two
units.  In a three generation model, the amplitudes is further
suppressed by mixing angles between the first and third generations, and
the total change in baryon number in the minimal instanton process is
three.

The zero modes of the heavy fields have consequences even within the
sector of zero topological charge, when we restrict attention to Green's
functions containing only the fields of light particles.
  Indeed, the heavy fermion
determinant in the presence of an instanton-antiinstanton pair factors
into the product of the determinants in each individual configuration
when the separation between the pair is large.  Since the instanton and
antiinstanton determinants vanish, the determinant in the pair
configuration must go to zero as the separation goes to infinity.  We have
noted above that the zero mode wave functions die exponentially.  The
pair determinant is thus an exponentially vanishing function of
separation.   In terms of the statistical mechanics of the dilute
instanton gas, this is equivalent to an attractive linear confining
potential between instantons and anti-instantons:
\eqn\instforce{Z_{I + \bar{I}} \sim \int d^4 R_I d^4 R_{\bar{I}}
 e^{- N m_F |R_I - R_{\bar{I}}|}}
where N is the number of heavy fermion zero modes.  Again we see that the
 dynamics of the low energy gauge fields is drastically affected by the
 virtual modes of the heavy mirror fermions.

Is there any kind of effective low energy
field theoretic description of the system we have studied at energies of
order $1$ GeV?  Certainly the conventional description, in which only
fields for the light particles are included, is wrong.  Recently a class
of models was described in which the low energy effective theory had to
be supplemented by a number of discrete global
variables.\ref\irrational{T. Banks, M. Dine,
N. Seiberg,{\it Phys.Lett.}{\bf B273},(1991),105. }\foot{Note that the
irrational couplings
which were the focus of\irrational are not necessary to the existence of
these global variables.  They exist in many perfectly renormalizable
four dimensional field theories.}  The resulting effective theory
violates the clustering axiom.  Is a similar description of decoupled
mirror fermions available?  We suspect that the answer is no.  In the
nonperturbative regime, the number of heavy fermion field modes which
are important to the low lying dynamics depends crucially on the
configuration of low energy boson fields.  Since the separation between
heavy and light fermion degrees of freedom is light field dependent, one
should not expect a local effective Lagrangian, unless we keep the
fields of the heavy particles in the low energy effective
Lagrangian.  There is no way that a local effective Lagrangian for the
light fields can produce a linear confining force between
instantons.\foot{This situation
bears a certain resemblance to that which occurs in theories
which have large numbers of degenerate, physically inequivalent vacuum
states.  In such theories, it is possible for a particle that is massive
at a generic point in the vacuum manifold to become massless at certain
special points.  The effective action obtained by integrating out this
massive particle at a generic point, becomes singular and non local at
the special points.  The new observation that we are making here is that
these nonlocal effects are also important for field configurations which
visit special points in the field manifold in a local region of spacetime.}

If the heavy fermion masses could really be taken to infinity we would
have a somewhat paradoxical situation in which the low energy theory
contained fields which created no particle states from the vacuum.  A
less radical description is suggested by the work of references\df
and\bag :
soliton states of the combined heavy fermion - Higgs boson system
survive at low energy even when the elementary fermion masses go to
infinity. These solitons have the quantum numbers of the elementary
fermions and masses of order the vacuum expectation value of the Higgs
field.  D'Hoker and Farhi were not able to firmly establish the existence of
such solitons, because they dealt with a theory of fixed length Higgs
fields and relied on the topology of the compact Higgs manifold as well
as on hypothetical short distance corrections to the effective action
that could stabilize the soliton configurations of the nonlinear model.
However,the existence of such solitons is also suggested by early work on
baglike nontopological
solitons\bag .   In
these references, the crucial ingredient is
the variable radius of the Higgs field. When a fermion gets its mass from a
Yukawa coupling, a single fermion state can exist in which the
value of the Higgs field vanishes near the location of the fermion.
Fermion modes of low energy exist in which the elementary fermion wave
function is trapped in the vicinity of this zero of the Higgs field,
avoiding the region of space where the fermion mass is large.  If
the energy required to locally deform the value of the Higgs field away
from its vacuum value is much less than the elementary fermion mass,
a light soliton state with elementary fermion quantum numbers is formed.
It is plausible that such states exist in the present model, though we
have not investigated the question in detail.  If they do, the necessity
of keeping the heavy fermion fields in the low energy Lagrangian would
no longer be paradoxical or bizarre.  They would be necessary to a
description of the light soliton states.

We should point out that in the model which we have described in this
section there is no really tight argument that a purely low energy
description of instanton processes should exist.  In a conventional
theory with a heavy sector, an effective local theory of the light
particles is supposed to describe low energy physics up to the accuracy
$({M_{light}\over M_{heavy}})^p$ for all positive $p$.  In our model
$ {M_{light}\over M_{heavy}} \sim {e\over g}$ where $e$ is the gauge
coupling, and $g$ the Yukawa coupling of the heavy fermions.  $g$ is
required to be of order $1$, so these effects are small only when $e$ is
very small.  On the other hand, the 't Hooft process is parametrically
of order $e^{- {8\pi^2 \over e^2}}$.  Thus it is smaller than effects of
the heavy particles, and we do not have the right to insist that it is
described correctly in terms of a low energy lagrangian.
(On the other hand, in trying to construct a
lattice standard model, we really want the lattice fermion doubles to go
off to infinite energy, leaving no trace behind.  In this case it is
crucial that there be a low energy Lagrangian which correctly describes
the symmetries of the model.)  We do not believe that this criticism of
our analysis is truly substantive.  The fermion zero modes and almost
zero modes that were
crucial to us exist for all values of the Yukawa coupling, and for all
configurations in the functional integral that have widely separated
lumps of topological charge.  There is no
indication that anything qualitatively new
happens when the Yukawa coupling begins to
leave the perturbative regime, other than the fact that the confining
force between instanton and anti-instanton gets stronger.

We are however deeply disturbed by the potential instabilities of the
perturbative vacuum in this model.
Thus, in order to confirm and enhance our understanding of
the picture of fermion decoupling that we have
presented here, we turn in the next section to
some two dimensional models.

\newsec{\bf Two Dimensional Models}

Our analysis of the weakly coupled mirror standard model suggests that
the failure of decoupling of mirror fermions is related to the existence of
configurations in which the Higgs field is equal to zero (or is at least
very small) at some point
in spacetime.  Indeed, the 't Hooft constrained instanton solution has
vanishing Higgs field at the core of the instanton, and the
mechanism for constructing light fermionic bag states also depends
on zeroes of the Higgs field.  If we restrict
attention to configurations in which the magnitude of the
Higgs field is everywhere
bounded from below by a positive constant $m_0$, it is probably possible
to use the methods of Witten and Vafa\ref\vitvaf{E.Witten,C.Vafa,{\it
Nucl. Phys.}{\bf B234},(1984),173.}
to prove that the effective action obtained by integrating over heavy
fermions contains nonlocality only over some finite scale of order
${1\over g m_0}$.  To confirm this intuition, we will study a two
dimensional model with fixed magnitude Higgs field.

The model that we will study has a $U(1)$ gauge symmetry.  It
contains two massless left moving fermions $\psi_i$, with gauge charges
$q_i$, and a massless right mover $\psi$ with charge $q$.  The charges
satisfy the anomaly cancellation condition $q^2 = q_1^2 + q_2^2$. We
will also include gauge singlet partners for each of these particles, in
order to describe them in terms of two component Dirac fields. We use
the letter $\Psi$ to denote the triplet of light Dirac fields. In
addition, we have mirror fermions $P_i$ and $P$ which are
rightmovers (resp. leftmovers) and carry charge equal to that of their
mirror partners.  The mirror fermions also have singlet partners, and we
will include Yukawa couplings between mirror fermions and Higgs fields
which provide Dirac masses for the mirror fermions in the presence of a
Higgs field vacuum expectation value.  The triplet of heavy Dirac fields
is denoted $P$.
The Lagrangian is
\eqn\twodlag{
\eqalign{
{\cal L} = & {-1 \over 4 e^2} F_{\mu\nu}^2
+ |\partial_{\mu} - i A_{\mu}\phi |^2 - \lambda V(\phi^{\dagger}\phi )
+ {\bar\Psi}i\gamma^{\mu}\bigl{[}\partial_{\mu}
- q\bigl{(}{1+\epsilon\gamma_3 \over 2}\bigr{)}A_{\mu}\bigr{]}\Psi \cr
+ & {\bar P}\Bigl{[}i\gamma^{\mu}\bigl{[}\partial_{\mu}
- q\bigl{(}{1-\epsilon\gamma_3 \over 2}\bigr{)}A_{\mu}\bigr{]}
+ g (\phi^{\dagger} )^q \bigl{(}{1 -\epsilon\gamma_3 \over 2}\bigr{)}
+ g (\phi )^q \bigl{(}{1+\epsilon\gamma_3 \over 2}\bigr{)}\Bigr{]} P \cr
}}
Here $q$ and $\epsilon$ are $3$ x $3$ matrices: $q = diag (q_1 , q_2 ,
q)$ and $\epsilon = diag (\epsilon_1 ,\epsilon_2 , \epsilon_3 ) = diag
(1, 1, -1)$.
The gauge coupling $e$ and the Yukawa coupling $g$ both have dimensions
of mass, and we take $g\gg e$.  The coefficient $\lambda$ in the Higgs
potential has dimensions of mass squared and determines the
spacetime scale of fluctuations of the radial mode of the Higgs
field.  We will first take this scale to be much larger than the Yukawa
coupling so that this mode does not participate in the physics at any
scale of interest.  Thus, the radial mode of the Higgs field is frozen:
$\phi = e^{i\theta}$.
Instanton configurations of the gauge-Higgs system will then have infinite
action.

In this limit it is convenient to tranform the heavy fermion fields by
multiplying them by functions of the Higgs fields in such a way as to
make them gauge invariant.
Let $ {\cal P} = e^{-iq \bigl{(}{1-\epsilon\gamma_3
\over 2}\bigr{)}\theta} P$ be the gauge transformed field; then the
Lagrangian becomes:
\eqn\unlag{
\eqalign{
 {\cal L} & =   {-1 \over 4 e^2} F_{\mu\nu}^2 + (\partial_{\mu}\theta -
A_{\mu})^2 +  {\bar\Psi}i\gamma^{\mu}\bigl{[}\partial_{\mu} -
q\bigl{(} {1+\epsilon\gamma_3 \over 2} \bigr{)}A_{\mu}\bigr{]}\Psi \cr
+ & {\bar {\cal P}} \bigl{[} i\gamma^{\mu}
[\partial_{\mu} - q\bigl{(}{1-\epsilon\gamma_3
 \over 2}\bigr{)}(A_{\mu} - \partial_{\mu}\theta )]
+ g \bigr{]}{\cal P} \cr }}
Note that the mirror fermions now have a constant mass term.  In the limit that
the gauge coupling becomes very weak (i.e. is much smaller than the
fermion mass) the model evidently reduces to a current current coupling
between a massive fermion and a massless Goldstone boson, a
renormalizable Lagrangian.  The coupling to the gauge boson is
superrenormalizable.

Let us now imagine doing a renomalization group analysis of the theory,
integrating out degrees of freedom above some cutoff scale $\Lambda$
which is much larger than $e$ and much smaller than the fermion mass.
In this integration, the gauge coupling can be treated perturbatively
since it is superrenormalizable and the fluctuating degrees of freedom
have an infrared cutoff.  The result of this integration is an effective
field theory for the light bosonic degrees of freedom $\phi$ and
$A_{\mu}$, and the massless fermions. The effective action depends only
on the gauge invariant
field $B_{\mu} = \partial_{\mu}\theta -  A_{\mu}$, which couples to a
chiral current of the heavy fermions.  We can classify the possible
terms in this effective action according to their dimension.  The only
term of dimension 2 is quadratic in $B_{\mu}$.  Its coefficient will be
logarithmically divergent in the limit $m\rightarrow\infty$\foot{In
fact, in the lowest order in the loop expansion this divergence cancels
when the contributions of all heavy fermion loops are summed.  It is
proportional to the anomaly.  This cancellation does not persist in
higher orders.}.  All other terms in the action have dimension greater
than two and their coefficients vanish in the heavy fermion limit. To
lowest order in $e$ the effective action is obtained from that of the
field $\theta$ in the theory:
\eqn\thirlag{
{\bar{\Psi} i\gamma^{\mu}(\partial_{\mu} -
q\partial_{\mu}\theta {(1 + \gamma_3)\over 2} )\Psi  + {m_V^2 \over 2 e^2}
(\partial_{\mu}\theta )^2},}
by the substitution $\partial_{\mu}\theta \rightarrow B_{\mu}$.  Higher
order corrections in $e$ vanish in the limit of large mass.

Although the above analysis is motivated by an examination of Feynman
diagrams we believe that it is valid nonperturbatively.  When $e=0$ the
model from which we obtain the effective action is a version of the two
species Thirring model with one of the fermions made extremely massive.
There seems to be no place for unexpected surprises.  If this is the
case, the decoupling of mirror fermions seems to work in this model.
Their effect on the low energy effective action is simply to introduce
an infinite renormalization of the gauge boson mass term $B_{\mu}^2$.
If we are willing to tune parameters to ensure that the gauge boson
remains light, then we obtain a chiral gauge theory in the limit
$m\rightarrow\infty$.

Not surprisingly, in this theory with fixed length Higgs fields, the
D'Hoker Farhi analysis of the baryon number current goes through.  If we
couple an external gauge field $a_{\mu}$ to the nonanomalous sum of
ordinary and mirror baryon number, it is easy to verify that in the low
energy theory $a_{\mu}$ couples to $J^{\mu}_L - (q_1 + q_2 -
q)\epsilon^{\mu\nu} (\partial_{\nu}\theta - A_{\nu})$.
However, we can also verify that in this model the result of this low
energy identification is to rule out the 't Hooft process in low energy
physics. As in section 2, the D'Hoker Farhi identification of the
divergence of the baryon number current with the divergence of a current
constructed from gauge invariant, massive, fields, suggests that global
baryon number is conserved.  In the present model the low energy theory
is exactly soluble (when the massless fermions are bosonized the
Lagrangian becomes quadratic), and we can verify this conjecture
explicitly. The simplest way to see this is to integrate out the massive
vectors, to obtain a baryon number conserving action for the massless
fermions:
\eqn\effact{
{\bar\Psi}i\gamma^{\mu}\partial_{\mu}\Psi + C \int dx dy{
J^{\mu}(x)\bigl{[}{g_{\mu\nu} + {\partial_{\mu}\partial_{\nu} \over M_V^2}
\over
-\partial^2 + M_V^2}\bigr{]}(x,y) J^{\nu} (y)}  }
where $J_{\mu} = {\bar\Psi}\gamma_{\mu} q \bigl{(}{1+\epsilon\gamma_3
 \over 2}\bigr{)}\Psi$.
 Although this action is nonlocal on the scale of the vector
boson Compton wavelength, it contains no infrared divergences, and no
violation of baryon number.

For completeness, we record the bosonized
form of the low energy action before the vector bosons are integrated
out.  Each low energy Dirac fermion is realized in terms of a scalar
field whose gradient is the {\it vector current} of the fermions. We
call the scalar corresponding to $\psi_i$, $\varphi_i$.
The bare Lagrangian is

\eqn\boslag{{\cal L} = {-1 \over 4 e^2} F_{\mu\nu}^2 +
(\partial_{\mu}\varphi_i - q_i A_{\mu} )^2 +
 (\partial_{\mu}\theta - A_{\mu})^2
+ \sum q_i \epsilon_i \varphi_i\epsilon^{\mu\nu}F_{\mu\nu}  }
After integrating out the heavy fields and rewriting things in terms of
the gauge invariant massive vector boson field $B_{\mu}$ , this becomes:
\eqn\boslagb{
{\cal L} = {-1 \over 4 e^2} B_{\mu\nu}^2 +
(\partial_{\mu}\Phi_i - q_i B_{\mu} )^2 +
 \alpha (B_{\mu})^2
+ \sum q_i \epsilon_i \Phi_i \epsilon^{\mu\nu}B_{\mu\nu}
}
where $\Phi_i = \varphi_i - q_i \theta$.  Note that we have had to use
the anomaly cancellation condition to show that $\theta$ does not appear
on this final form of the Lagrangian.
The fixed length Higgs model described above thus realizes the goal
that one would like to achieve in constructing the standard model on the
lattice.  However, it does so at the expense of eliminating the 't Hooft
process from low energy physics.  This is perfectly consistent within
the framework of the low energy effective Lagrangian, where the 't Hooft
amplitudes clearly vanish in the limit that the mass of the Higgs boson
goes to infinity.\foot{In four dimensional non Abelian gauge theories,
 the validity of this claim is
not obvious, although it is certain that the conventional instanton
action becomes infinite with the Higgs mass.  It is hard to discuss the
question rigorously, since the entire theory becomes strongly coupled as
the Higgs mass gets large, and probably the limiting theory does not
exist.\dashneuberg}
In four dimensions we do not know of a consistent version of the
standard model with an arbitrarily heavy Higgs particle, so the above
scenario cannot be achieved.

In order to study a two dimensional model
with variable length Higgs in a reliable manner, we introduce $N$ copies
of both the low energy and mirror fermions, and take the limit
$N\rightarrow\infty$ with $e^2 N = E^2 $, $g^2 N = G^2$, and
$\lambda N = \kappa$ fixed.  In this limit, quantum fluctuations of the
boson fields are suppressed, while the ratio of tree level gauge boson
to fermion masses is $E\over G$, and can be as small as
we like.

To leading order in $N$, the theory is solved by finding stationary
points of the effective action:
\eqn\seff{
\eqalign{
S_{eff} = & N[  {| D_{\mu}\phi |}^2 - \kappa (|\phi |^2 - 1)^2 -
{1\over 4 E^2} F_{\mu\nu}^2 + {\rm Tr\quad ln}[i\gamma^{\mu}(\partial_{\mu} -
q A_{\mu} ({1 + \epsilon\gamma_3 \over 2})] \cr
+ & {\rm Tr\quad ln}[i\gamma^{\mu}(\partial_{\mu}
- q A_{\mu} ({1 - \epsilon\gamma_3 \over 2}) - G {\phi^{\dagger}}^q({1 -
\epsilon\gamma_3 \over 2}) - G \phi^q ({1 + \epsilon\gamma_3 \over 2})] \cr
}}
The large $N$ vacuum state is determined by stationary points of this
effective action with vanishing gauge fields and constant Higgs fields.
The heavy fermion contribution to the effective potential for the Higgs
field dominates the classical term for $G^2 \gg \kappa$.  It has the
form:
\eqn\veff{V_{ferm}(\phi ) = {G^2 \over 4\pi} |\phi |^2 {\rm ln}(|\phi |^2 )}
This potential is shown in Fig. 1.  It has the typical spontaneous
breakdown form, and is bounded from below.  It determines the minimum of
the Higgs field to lie at $|\phi |^2 = e$, and the Higgs mass,
determined as the curvature of the potential at its minimum, is of order
$G^2$.  Although this seems to be a consistent theory, it is not what we
want if we intend to decouple the heavy fermions while keeping the Higgs
boson light.  In that case we
expect to keep the Higgs particle at low mass, and we may attempt to do
this by fine tuning the coefficients of relevant operators in the low
energy theory.  In two dimensions there are an infinite number of
relevant operators for a scalar field, although conventional
renormalization theory leads us to expect only a quadratic term in this
leading $N$ approximation.  In order to keep the minimum at its
classical value $\phi = 1$ and keep the Higgs mass of order
$\sqrt{\kappa}$, we need to tune at least two parameters.  The quartic and
quadratic couplings of the classical Lagrangian suffice, but the
resulting potential has a negative quartic coupling and is unbounded
from below.  The addition of a $|\phi |^6$ coupling allows us to keep
the potential bounded.  There is then one free parameter.  For all
values of this parameter, the resulting potential has a deeper minimum
either much closer to or much further from the origin than $\phi = 1$.
For a ratio of $100$ between the fermion and Higgs masses, the potential
typically looks like Fig. 2.  The perturbative vacuum with small Higgs
mass that we have constructed by fine tuning three parameters is
metastable and rather short lived.
One must add higher order terms to get sensible results.  After a while
it dawned on us that what we were doing could best be described as
follows:  {\it for any value of ${G^2 \over \kappa}$ find a polynomial
approximation $P(\phi )$ to $|\phi |^2 {\rm ln}(|\phi |^2 )$ which
approximates
this function with accuracy ${\kappa\over G^2} $ in a range $0 \leq
|\phi | \leq \phi_0$ with  $\phi_0 > 1$.  Arrange further that $V_{ferm}
(\phi ) - {G^2 \over 4\pi} P(\phi )$ be positive and monotonically
increasing for $\phi > \phi_0$.  Then add $- {G^2 \over 4\pi} P(\phi )$
to the classical potential.}  The resulting effective potential looks
just like the classical potential for $\phi < \phi_0$ and shoots up
dramatically beyond this point.  With sufficient fine tuning we can even make
$\phi_0$ very large.  As the fermion mass gets larger we need to tune
more and more parameters to obtain a low energy effective potential that
agrees with that in a theory where the fermions are absent.  A similar
situation would be found if we tried to use a basis of analytic
functions other than polynomials\foot{We might for example use operators
of fixed dimension at the Gaussian fixed point, i.e. sines and cosines.}
to construct our local counterterms.  We would still need a number of
parameters which grew with the fermion mass to construct a satisfactory
theory with a light Higgs field and a stable vacuum.

The reason for the difficulty we had in obtaining a satisfactory low
energy potential is not hard to find.  The fermion contribution to the
potential is not analytic at the origin.  This is a consequence of
infrared divergences which occur because the fermion is massless
when the Higgs field vanishes.  If the potential had been an entire
function, we could have cancelled it exactly with a sequence of allowed
counterterms.  This difficulty is familiar from four dimensions, and was
the origin of the instability of the perturbative vacuum in our
superweakly coupled standard model with heavy fermions and light Higgs
bosons.  The two dimensional example shows that the problem is more
general than the unboundedness of the fermion induced effective
potential, for in two dimensions that object is perfectly well behaved.
Rather, it is the attempt to make the scalar field whose VEV was
responsible for the fermion mass much lighter than the fermion itself
which was the cause of the problem.  In this was of saying things, it
becomes clear that these difficulties are not restricted to the
decoupling of fermions.  Indeed, the Linde-Weinberg \linde\ lower
bound on the Higgs boson mass may be viewed as an example of the same
phenomenon. Looked at from the point of view of an effective field
theory for the conjectural light Higgs boson, the two problems are
almost identical.  It is only because we have always viewed this problem
from the vantage point of the heavy scale (the gauge boson masses) that
it has not caused the same confusion.  The statement that the standard
model vacuum is not stable unless the Higgs boson mass is greater
than a certain finite fraction of the gauge boson mass is equivalent to
the statement that one cannot decouple the heavy gauge boson from an
effective field theory for the light Higgs boson, despite the fact that
the ratio of their masses at tree level appears arbitrary.  Again, the
problem is caused by the size and nonanalyticity of the effective Higgs
potential induced by the heavy particles.

It is also amusing to note that the local terms in the effective
Lagrangian which describe the failure of decoupling of chiral fermions
in perturbation theory (and in particular, the Peskin-Takeuchi S
parameter), are also nonanalytic at vanishing Higgs field.  When written
in a gauge invariant manner, they have the typical
Higgs dependence\foot{T.B. thanks L.
Randall for explaining this point to him.}:
\eqn\Spar{{H^{i_1} \ldots H^{i_n} \over |H|^n}}
This suggests that they may also be viewed as coming from infrared
divergences.

Suppose now that we have performed the massive fine tuning described
above and constructed a theory with a stable vacuum and a Higgs boson to
heavy fermion mass ratio which is very small.  To all orders in the
$1\over N$ expansion, the theory will conserve the baryon number of the
light fields.  To investigate whether this continues to be true
nonperturbatively in $N$ we look for solutions of the equations of
motion of the large $N$ effective action which carry non-zero
topological charge.  There are none! In any configuration of gauge and
Higgs fields with nonzero topological charge,
 the heavy fermions will have normalizable zero modes.  The fermion
 determinant vanishes, and the effective action of instantons is
 infinite.  Note that the polynomial potential $P(\phi )$ cannot
change this conclusion.  Like the fermion mass, it is finite but large.
It cannot cancel an infinity coming from the zero mode.

\lref\jackiw{R. Jackiw and P. Rossi, Nucl. Phys. {\bf B190} (1981) 681.}
\lref\weinberg{E. Weinberg, Phys. Rev. {\bf D24} (1981) 2669.}

We would now like to exhibit fermion zero modes
in the instanton background in a more explicit manner.
To this end, we study a single charged Dirac field in
the instanton background, with Lagrangian:
\eqn\lagrangian{
{\bar\psi} \gamma^\mu ( i \partial_\mu - e A_\mu
({{1 + \gamma_3}\over 2}) ) \psi
 - g^* \phi^* {\bar\psi}   ({{1 + \gamma_3}\over 2}) \psi
 - g    \phi  {\bar\psi}   ({{1 - \gamma_3}\over 2}) \psi    \qquad . }
The instanton configuration with winding number $n$ is given by
\eqn\instanton{
\eqalign{
e A_\mu  = & \epsilon _{\mu\nu} {\hat x}^\nu A (r) \cr
g \phi   = & i e^{-i n \theta} f(r) \cr
}}
At the core of the instanton $A(r) \sim 0 $ and $f(r) \sim
r^{|n|}$
whereas at infinity $A(r) \sim +n/r$ and $f(r) \sim {\rm constant}$.
Here we have chosen to work in Landau gauge.
The Dirac equation in this background is similar to the one analyzed in
\jackiw\ for a fermion-vortex
system where zero modes of definite chirality
were guaranteed by an index theorem \weinberg\ .
In our problem the  Higgs coupling
is slightly different and there are no chiral zero modes.
We will therefore explicitly solve the equation to find
$n$ normalizable zero modes in this sector.
Substituting $$ \psi = \left(\matrix{P\cr
                                    Q\cr}\right) =
                       \left(\matrix{p e ^ {+\int A(r)} \cr
                                    q\cr}\right) $$
the Dirac equation becomes
\eqn\dirac{
\eqalign{
e^{i\theta} ({\partial \over \partial r} + {i\over r}{\partial \over
\partial \theta}) q = & -f(r) e ^ {+\int A(r)} e^{+i n \theta} p \cr
e^{-i\theta} ({\partial \over \partial r} - {i\over r}{\partial \over
\partial \theta}) p = & -f(r) e ^ {-\int A(r)} e^{-i n \theta} q   .  \cr
}}
To separate the angular dependence we write
$ p =  e^{-i m \theta } p_m(r) $ and
$ q =  e ^{i (n - m - 1)\theta} q_m(r) $ and obtain the following coupled
first order equations:
\eqn\radial{
\eqalign{
 ({\partial \over \partial r} - {m\over r}) p_m
=& -f(r) e ^ {\int A(r)}q_m \cr
 ({\partial \over \partial r} - {n-m-1\over r}) q_m
=& -f(r)  e ^ {-\int A(r)} p_m       . \cr
}}
We can turn these into a single second order differential equation for
either $p_m$ or $q_m$, which will have two linearly independent
solutions.
For large r, the fermion is massive; so apart from powers of $r$ the
two solutions go as $e^{\pm \mu r}$. Only $e^{-\mu r}$ is acceptable
as a normalizable solution.

At the origin, from \radial\ and
using the asymtotics
$f(r) \sim r^{|n|}$ and $ e ^ {-\int A(r)} \sim {\rm constant}$
we see
that the two solutions go as
\eqn\solution{
\eqalign{         p_m \sim r^m ,\, & q_m \sim r^{|n| + m+1} \cr
{\rm and }\qquad  p_m \sim r^{|n|+ n -m} ,\, & q_m \sim r^ {n-m-1}  . \cr
}}
In general, the solution that is well behaved at infinity will be an
arbitrary linear combination of these two solutions, which should also
be well behaved at the origin.
Thus, from \solution\ we see that, for positive $n$, there are $n$
normalizable zero energy
solutions for $0\leq m\leq n-1$. For negative $n$, the normalizable zero modes
come from the Dirac equation for ${\bar \psi}$ (Note that in Euclidean
space, $\psi$ and ${\bar \psi}$ are not related
by complex conjugation.)  As a consequence, the `t Hooft effective
action will violate the fermion number symmetry of these heavy fermions,
as required by the anomaly, leading to the physical effects described above.

In particular, as noted above, the large N effective action will not have
finite action
instanton solutions with nonzero topological charge.
If we consider configurations of zero topological charge which consist
of two widely separated lumps of charge of opposite sign, then, as in
four dimensions, the fermion determinant will contribute an effective
confining force between instantons and anti-instantons.   The confinement of
instantons leads to another dramatic effect, which can be studied
without the aid of light chiral fermions.  The purely bosonic abelian
Higgs model exhibits confinement of external charges which are fractions
of the charge on the Higgs field.  The mechanism for confinement is a
dilute gas of instantons.
We now see that the \lq\lq mere''
introduction of very heavy chiral fermions into the theory completely
eliminates this nonperturbative and nonlocal effect.  The heavy chiral
fermions do not decouple as their mass goes to infinity.  In this
context it is even more apparent that local counterms cannot mimic these
effects. The appearance of a confining force between instantons and the
corresponding disappearance of the confining force between fractional
charges will not be affected by the inclusion of local gauge invariant
terms in the bosonic action.

\newsec{Applications to Lattice Gauge Theories}

The results that we have obtained for continuum models suggest analogous
problems in any lattice gauge theory which attempts to decouple lattice
fermion doubles by using the device of a Wilson-Yukawa coupling to a
Higgs field.  This includes all of the models studied in
\ref\smit{
J. Smit, Talk presented at Int. Conf. 'Lattice '89, Capri, Italy, Sep
18-21, 1989,
{\it  Capri,Lattice 1989}, 3; W. Bock, A.K. De, K. Jansen, J. Jersak,
T. Neuhaus, J. Smit, {\it Phys.Lett.} {\bf B232} (1989) 486;
P.V.D. Swift, {\it Phys.Lett.} {\bf 145B} (1984) 256;
M.F.L. Golterman, D.N. Petcher, WASH-U-HEP-91-60; WUSL-HEP-91-GP1;
FSU-SCRI-90C-185, Talk presented at Lattice '90 Conf., Tallahassee, FL,
Oct 8-12, 1990; S. Sanielevici , H. Gausterer,
M.F.L. Golterman, D.N. Petcher, FSU-SCRI-90C-183, Nov 1990,
to appear in Proc. of Lattice '90 Conf., Tallahassee, FL, Oct 8-12,
1990; M.F.L. Golterman, D.N. Petcher {\it Nucl.Phys.} {\bf B359} (1991) 91;
{\it Phys.Lett.} {\bf B247} (1990) 370;
{\it Phys.Lett.} {\bf B225} (1989) 159; S. Aoki, I.H. Lee, J.Shigemitsu,
R.E. Shrock,
{\it Phys.Lett.} {\bf B243} (1990) 403;
S. Aoki, I.H. Lee, D.Mustaki, J. Shigemitsu, R.E. Shrock,
{\it Phys.Lett.} {\bf B244} (1990) 301.}.

Strictly speaking, our analysis applies only in the spontaneously broken
phase of the theory.  Lattice analysis had already led to the
conclusion that the Wilson-Yukawa method does not work in this phase.
Much analysis has been devoted to the symmetric phase of these
models.  When the Higgs field in the symmetric phase is allowed to have
a mass of the order of the cutoff, we can achieve a symmetric phase in
which the absolute value of the Higgs field is not small.  Symmetry is
achieved by making local quantum singlet states by superposing states
with the same large magnitude of the Higgs field, but different
orientations in group space.  There can be no analog of this phase for a
continuum Higgs field. The Higgs bilinear which appears in the
Wilson-Yukawa coupling is not small in such a phase, and this term in
the action can provide a mass to fermion doubles.  However all attempts
to utilize this mechanism to construct chiral gauge theories have
failed. The fermions always appear in vector representations of the
gauge group \ref\petcher{M.F.L.Goltermann, D.N.Petcher,J.Smit,
{\it Nucl. Phys.} {\bf B370},(1992),51.}.
\vskip.25em\hskip.5em
With a bit of hindsight and a bit of effective field theory, we can
understand why this failure was inevitable.  As usual in theories with
Wilson terms one must perform fine tuning in order to make some of the
fermions in the theory massless.  This means that the erstwhile chiral
gauge theory is part of a continuum of theories in which the masses of
the massless fermions are nonzero but very small on the scale of the
lattice spacing.  Now consider an effective field theory for these
light, but not exactly massless, fermions.  It must be a gauge theory
with no spontaneous breakdown, for we are in the symmetric phase.  But
it must also contain mass terms for the light fermions.  This means that
the light fermions can have gauge invariant masses, and are thus in
vector representations of the gauge group.

The only lattice gauge theories which can avoid the problems we have
described are those which do not use a Higgs field to decouple the
fermion doubles.  These fall into two categories.  The Rome
approach\ref\rome{L. Maiani, G.C. Rossi, M. Testa, {\it Phys.Lett.}{\bf
B261},(1991),479;
A. Borrelli, L. Maiani, R. Sisto, G.C. Rossi, M. Testa,{\it
Nucl.Phys.}{\bf B333},(1990),335;
M. Bochicchio, L. Maiani, G.
Martinelli, G. C. Rossi, M. Testa,{\it Nucl.Phys.}{\bf
B262},(1985),331.}
puts a gauge fixed theory on the lattice.  Non gauge
invariant Wilson terms are added to decouple the doubles, as well as a
host of nongauge invariant counterterms whose coefficients are supposed
to be fine tuned to achieve BRST invariance in the continuum.  As a
consequence of the explicit choice of gauge, the theory is not equivalent
to a gauge invariant lattice theory with a Higgs field.\foot{A formal
argument seems to show that the equivalence is reinstated in the
continuum limit, but this argument neglects wildly fluctuating lattice
Higgs modes.  This subtle point was explained in great detail by J.
Smit, M. Golterman, D. Petcher, H. Neuberger, L. Maiani, and M. Testa in
informal discussions at the Rome conference on chiral lattice gauge theories.}
In particular, Dugan and Randall\ref\dugan{
ON THE DECOUPLING OF DOUBLER FERMIONS IN THE LATTICE STANDARD MODEL.
By M. J. Dugan, L. Randall, MIT preprint, MIT-CTP-2050, Jan 1992.
{\it Submitted to Nucl.Phys.B}.} have shown that the fermion
doubles do
not lead to a contribution to the Peskin-Takeuchi S parameter in this
model.  When applied to the lattice standard model, this approach
appears to contain an unwanted baryon number symmetry that the continuum
model does not have.  L. Maiani has argued that the current of this
symmetry may not be BRST invariant in the continuum limit.  It may
indeed be correct that this is the meaning of the Dugan-Manohar
calculation in the context of the Rome model (we have argued that it has
quite a different meaning in the Swift-Smit model).  Nonetheless, we
remain disturbed by the fact that within this model we cannot write down
a lattice Green's function which violates baryon number.
In order for Maiani's argument to completely resolve the baryon number
paradox in this model, we must demonstrate that the bothersome $U(1)$
symmetry is spontaneously broken on the lattice.  Maiani's argument
could then be used to show that the corresponding Goldstone boson was an
unphysical gauge excitation.  It seems that a lot of work must be done
to prove that the Rome approach can really reproduce the continuum
standard model. In
applications to strongly coupled chiral gauge theories like the SU(5)
model, there does not seem to be a similar problem with the Rome
approach.

We note also that serious questions about the treatment of Gribov ambiguities
have been raised in connection with this approach.  In addition
Parisi\ref\parisi{G. Parisi, {\it Summary Talk at the Conference on
Chiral Gauge Theories}, Rome, March 9-13.  See the Conference
Proceedings.} has made the very interesting suggestion that conventional
renormalization group arguments about the relevance of operators which
break a gauge symmetry may fail in the presence of gauge field
configurations belonging to nontrivial fiber bundles.  The corresponding
vector potentials are singular somewhere in spacetime (perhaps at
infinity) and naive power counting arguments may not be applicable.  We
do not know whether either of these two potential problems with the Rome
approach is real.

The only model of which we are aware that escapes completely from the
problems that we have described, is the staggered fermion model
of\ref\smit{J. Smit,  ATTEMPTS TO PUT THE STANDARD MODEL ON THE LATTICE,
Contribution to Conf. LAT 87, Seillac, France, Sep 28 - Oct 2, 1987.
Published in  Seillac Sympos.1987:451}.
This model has no extra fermion degrees of freedom on
the lattice; the doubled modes are identified with known continuum
fermions.  The only consistent way to do this is to break color SU(3) on
the lattice, or equivalently to introduce colored Higgs fields.  This we
consider a point in the model's favor, for it destroys the baryon number
symmetry which was the source of all of our worries.  It remains to be
seen whether enough tuning of parameters can be done in this model to
truly reproduce the standard model, but we see no obvious
reason for it to fail.

Finally, we should mention the model of Eichten and
Preskill\ref\preskill{E.Eichten, J. Preskill, {\it Nucl. Phys.}{\bf
B268},(1986),179.}.  Recently Golterman and
Petcher\ref\ep{M.F.L.Golterman,D.Petcher, ON THE EICHTEN-PRESKILL
PROPOSAL FOR LATTICE CHIRAL GAUGE THEORIES, Washington University
preprint, WASHU-HEP-91-61.} have suggested that this suffers from the
same problems as the Smit-Swift models, despite its careful attempt to
break all unwanted symmetries by adding multifermion terms to the
lattice Lagrangian.  We do not understand the physics of either the
original model or the recent criticism of it very well.  If the
criticism is incorrect, the Eichten Preskill model may also provide a
convenient method for simulating the standard model.

\newsec{Conclusions}

We have demonstrated fairly conclusively that the super weakly coupled mirror
standard model introduced in section I, has a low energy sector whose
nonperturbative physics is not described correctly by a local Lagrangian
for the fields of the light particles of the tree level analysis.  This
result is confirmed in the two dimensional model with soft Higgs fields.
There we were able to make the mass ratio between the heavy fermions and
gauge bosons arbitrarily large by letting the number of fermion
multiplets tend to infinity.  The zero modes of massive fermions in
instanton fields
showed up directly as a contribution to the large $N$ effective action.
We also studied the limit of rigid Higgs fields in the two dimensional
model, and showed that although the heavy fermion fields in this model
truly decoupled, the low energy theory had no baryon number violation.

It seems to us that phenomena analogous to those we have described would
afflict any
lattice version of the standard model
 with an exactly conserved baryon number current,{\it if
one succeeded in eliminating all unconventional particles from the
continuum spectrum.}  By analogy with the model studied here, one would
suspect that no good continuum limit with such a spectrum could exist,
and if a limit were to exist it would certainly not be the standard model.
Explicit baryon number violation must be incorporated into lattice
versions of the standard model if they are to converge to the right
answer.  We caution that it is by no means certain that this necessary
condition is a sufficient one. If the conjectures that have been made about
Lee-Wick/SLAC/D'Hoker-Farhi
solitons are correct, then one might expect light states
with fermion quantum numbers to exist in almost any theory in which
fermion masses come solely from the Higgs mechanism.  In the limit of
large Yukawa coupling, the masses of these states are determined
primarily by the dynamics of the Higgs field.  Only by sending the
renormalized Higgs mass to infinity can we expect to decouple these
soliton states.  In the two dimensional model we found that baryon
number violation also vanishes in this limit, in accord with the general
argument that in this limit the anomaly is the divergence of a gauge
invariant massive operator.

In four dimensions, it seems unlikely that there will be a sensible
continuum limit for any spontaneously broken nonabelian gauge theory
with an infinite Higgs mass.  Thus, lattice
models with Wilson-Yukawa terms cannot reproduce the spectrum of the
standard model. In the previous section
we described the class of extant lattice models which may evade this
conclusion.  Our results also suggest the extra fermions which
transform chirally under the standard model gauge group will be found at
scales not too far removed from the weak scale, if they exist at all.
The question of the chiral nature of the weak interactions should be
settled once and for all by the next generation of accelerators.
\vfill\eject
\centerline{\bf Acknowledgements}
We would like to thank our colleagues N. Seiberg and H. Neuberger for
interesting conversations.  T.B. would also like to thank M. Goltermann,
D. Petcher, L. Maiani,
J. Smit, L. Randall, J. Shigemitsu, R. Shrock and S. Aoki
for discussions of lattice decoupling.  This work was supported in part
by DOE contract DE-FG05-90ER40559
\listrefs
\end